\documentclass[aip,rsi,superscriptaddress,amsmath,amssymb,reprint,longbibliography,floatfix,nofootinbib]{revtex4-2}
\usepackage{siunitx}
\usepackage{graphicx}
\usepackage{placeins}
\usepackage{lettrine}
\usepackage{adjustbox}
\usepackage{multirow}
\usepackage{xcolor}
\usepackage{bm}
\usepackage[colorlinks=true,
            linkcolor=blue,
            citecolor=blue,
            urlcolor=blue]{hyperref}

\begin{document}
\title{A nonclassical law of the wall in superfluid helium-4}

\author{Mikai Hulse}
\affiliation{National High Magnetic Field Laboratory, Tallahassee, FL 32310, USA}
\affiliation{Department of Physics, Florida State University, Tallahassee, FL 32310, USA}

\author{Musfequs Salehin}
\affiliation{National High Magnetic Field Laboratory, Tallahassee, FL 32310, USA}
\affiliation{Department of Mechanical and Aerospace Engineering, FAMU-FSU College of Engineering, Florida State University, Tallahassee, FL 32310, USA}

\author{Sosuke Inui}
\affiliation{National High Magnetic Field Laboratory, Tallahassee, FL 32310, USA}
\affiliation{Department of Mechanical and Aerospace Engineering, FAMU-FSU College of Engineering, Florida State University, Tallahassee, FL 32310, USA}

\author{Wei Guo}
\email[Corresponding: ]{wguo@fsu.edu}
\affiliation{National High Magnetic Field Laboratory, Tallahassee, FL 32310, USA}
\affiliation{Department of Mechanical and Aerospace Engineering, FAMU-FSU College of Engineering, Florida State University, Tallahassee, FL 32310, USA}


\date{\today}

\begin{abstract}
The logarithmic law of the wall is one of the most robust scaling laws of classical turbulence, yet whether it survives in a quantum fluid such as superfluid $^4$He (He II) remains unknown. At a solid wall, the viscous normal-fluid and inviscid superfluid components of He II obey fundamentally different boundary conditions, yet their motions can be coupled through quantized vortices. How these competing effects organize the near-wall flow remains an open question in quantum-fluid hydrodynamics. Here we visualize the near-wall velocity field in mechanically driven He II pipe flows at Reynolds numbers up to $1.26\times10^6$ using molecular tagging velocimetry and independently determine the wall shear stress from pressure-drop measurements. We find that the normal-fluid velocity develops a robust logarithmic profile, showing that a law-of-the-wall description persists despite the two-fluid nature of He II. Strikingly, however, the resulting wall law is distinctly nonclassical: the effective K\'arm\'an constant is less than half its classical value and remains nearly unchanged across the temperature range studied. This invariance, despite substantial changes in the relative fractions of the two components, indicates strong dynamical coupling between them. These results not only reveal how a universal law of classical turbulence is transformed by quantum hydrodynamics, but also provide valuable guidance for the design and operation of cryogenic cooling systems involving He II transport.
\end{abstract}
\maketitle

\section*{Wall turbulence in a quantum fluid}
The logarithmic law of the wall is one of the central organizing principles of classical turbulence \cite{vonKarman1930,Millikan1938,BradshawHuang1995,Smits2011,Marusic2013}. At sufficiently large Reynolds (Re) numbers, flows near smooth solid boundaries develop a logarithmic mean-velocity profile described by $u^{+}=\frac{1}{\kappa}\ln y^{+}+B$ \cite{Osterlund2000,Zagarola1998,McKeon2004Mean,Monty2009,LeeMoser2015,Marusic2013,Furuichi2015,Furuichi2018}. Here $u^{+}=u/u_{\tau}$ is the mean velocity normalized by the friction velocity $u_{\tau}$, while $y^{+}=yu_{\tau}/\nu$ is the dimensionless wall-normal distance, with $\nu$ denoting the kinematic viscosity. The friction velocity is defined by $u_{\tau}=(\tau_{w}/\rho)^{1/2}$, where $\tau_{w}$ is the wall shear stress and $\rho$ is the fluid density. The parameters $\kappa$ and $B$ are the K\'arm\'an constant and additive constant, respectively. Remarkably, $\kappa$ remains close to $0.4$ across a wide range of classical turbulent flows \cite{Osterlund2000,NagibChauhan2008,Smits2011,Marusic2013,Bailey2014,LeeMoser2015,Furuichi2015,Furuichi2018}, making the logarithmic law one of the most robust manifestations of universality in classical turbulence. However, for quantum fluids with fundamentally different boundary physics, whether this universal law survives remains an outstanding open question.

Superfluid $^4$He (He II) provides a particularly useful system to address this question. Below the lambda transition, He II behaves as two interpenetrating fluids: a viscous normal fluid and an inviscid superfluid \cite{Landau1941,Khalatnikov1965}. Rotational motion of the superfluid is confined to quantized vortex lines, each carrying a fixed circulation of $10^{-7}$~m$^2$/s \cite{Donnelly1991}. At finite temperature, interactions between the vortices and the normal fluid produce mutual friction between the two fluids, which can couple their motions \cite{HallVinen1956,Vinen1957}. At a solid wall, however, the two fluids obey fundamentally different boundary conditions: the normal fluid satisfies the conventional no-slip condition, whereas the superfluid is not subject to viscous no slip \cite{Landau1941,Khalatnikov1965}. The near-wall flow must therefore emerge from the interplay between these distinct boundary responses and vortex-mediated coupling between the two fluids.

Previous numerical studies have considered the behavior of He II near solid boundaries \cite{GodfreySamuels2000,YuiFujimotoTsubota2015,Galantucci2015,Stagg2017,Doyle2024}, but none has addressed the universal scaling regime in high Reynolds number He II flows. A useful simulation in this regime would require simultaneously resolving high-Re turbulent normal-fluid motion, a dense three-dimensional tangle of quantized vortices, and their two-way coupling across widely separated length scales\cite{Kivotides2000Triple, Yui2020Fully, Galantucci2020EPJ}, which is computationally formidable. Direct experiments therefore offer the most practical route to gain insight. Xu and Van Sciver previously used particle image velocimetry to measure velocity profiles in He II pipe flows and observed a turbulent boundary layer broadly consistent with classical correlations~\cite{Xu2007}. However, their measurements were limited to $Re$ of order $10^4$-$10^5$ and lacked sufficient near-wall resolution to examine the logarithmic law of the wall.

\begin{figure*}[t]
\centering
\includegraphics[width=\linewidth]{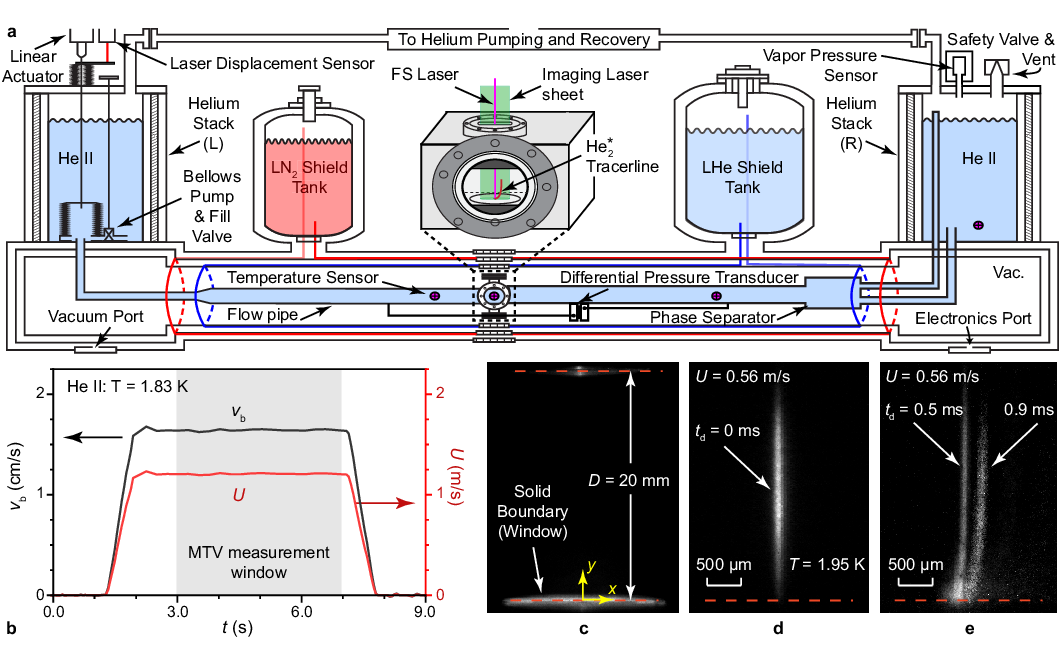}
\caption{\textbf{Experimental platform for near-wall velocimetry in He II.}
\textbf{a} Schematic of the Liquid Helium Flow Tunnel (LHeFT). A bellows pump in the left helium stack displaces He II through the $20\times20$~mm$^2$ square flow pipe toward the opposite stack. The measurement section provides optical access for molecular tagging velocimetry (MTV), and two pressure taps separated along the pipe are used for differential-pressure measurements.
\textbf{b} Representative bellows velocity $v_b$ and the corresponding mean velocity $U$ in the pipe. The shaded region indicates the MTV measurement window.
\textbf{c} Boundary image used to locate the solid walls for length scale calibration.
\textbf{d} Representative image of the He$_2^*$ tracer line immediately after its creation.
\textbf{e} Representative superimposed images of tracer lines recorded at two different drift times $t_d$ under a mean He II flow speed of $U=0.56$~m/s. Each tracer-line image shown in \textbf{d} and \textbf{e} is formed by accumulating 2000 single-shot exposures.}
\label{Fig1}
\end{figure*}

A reliable experimental characterization of the near-wall flow requires both a spatially resolved velocity profile and an independently determined wall shear stress. We achieve this by combining molecular tagging velocimetry (MTV) with pressure-drop measurements in mechanically driven He II pipe flows~\cite{Sanavandi2020}. By tracking thin lines of metastable He$_2^*$ molecular tracers~\cite{Guo2009,Gao2015,Marakov2015}, we quantitatively measure the near-wall normal-fluid velocity field. To determine the wall shear stress, we use its relation to the Darcy friction factor $f_D$: $\tau_w=\rho_\mathrm{He}U^2f_D/8$~\cite{McKeon2004Friction}, where $U$ is the cross-sectionally averaged flow velocity and $\rho_\mathrm{He}$ is the total density of He II. We determine $f_D$ from the pressure drop $\Delta P$ measured over a streamwise distance $L_p$ through $f_D=2\Delta P D_h/(\rho_\mathrm{He}L_pU^2)$~\cite{McKeon2004Friction}, where $D_h$ is the hydraulic diameter of the pipe. Thus, the pressure-drop $\Delta P$ provides $f_D$ and, in turn, an independent determination of $\tau_w$ for normalizing the measured velocity profiles. As we show below, He II develops a logarithmic wall law despite the fundamentally different boundary physics of its two fluid components. However, its scaling is distinctly nonclassical, revealing an emergent near-wall state organized by vortex-mediated coupling between the two fluids. These results extend one of the defining structures of classical turbulence into the quantum regime.

\section*{Experimental setup and flow measurements}
The experiments are performed in the Liquid Helium Flow Tunnel (LHeFT), shown schematically in Fig.~\ref{Fig1}a \cite{Sanavandi2020}. This facility consists of a smooth, $3.35$-m-long square flow pipe with an inner side length $D=20$~mm (hence $D_h=D=20$~mm), connecting two vertical helium storage stacks. Liquid helium in the stacks and flow pipe can be cooled to the desired temperature by regulating the vapor pressure. Flow through the pipe is generated by a bellows pump installed in one stack and driven by a computer-controlled linear actuator. With an effective bellows area of $A_{\rm b}=294$~cm$^2$, the mean flow velocity is determined geometrically from the bellows velocity $v_{\rm b}$ as $U=(A_{\rm b}/D^2)v_{\rm b}=73.5v_{\rm b}$. The $9.4$-cm bellows stroke displaces up to about $2.7$~L of liquid helium, producing a finite-duration flow with a well-defined steady interval. A representative bellows-velocity trace and the corresponding mean flow velocity are shown in Fig.~\ref{Fig1}b.

The near-wall velocity is measured using MTV with metastable He$_2^*$ excimer molecules \cite{Gao2015}. As illustrated in Fig.~\ref{Fig1}a, the measurement section is equipped with vertically aligned optical windows for the tagging and imaging laser beams and a front window for fluorescence imaging. The $24$-mm-diameter front window is wider than the inner width of the pipe, allowing the tracer line to be imaged continuously from the pipe interior to the solid boundary. The tagging optics can also position the tracer line at controlled distances from the wall to resolve the near-wall velocity variation~\cite{Sanavandi2020}. Before each MTV measurement, a boundary image is acquired under ambient illumination, as shown in Fig.~\ref{Fig1}c. The known $20$-mm pipe width provides the length-scale calibration, while the visible solid boundaries define the wall positions and hence the wall-normal coordinate $y$.

For each MTV acquisition, a focused $35$-fs tagging laser pulse at $780$~nm creates a thin line of metastable He$_2^*$ triplet molecules by field ionization of $^4$He atoms \cite{Gao2015}. These molecules have a lifetime of about $13$~s in liquid helium \cite{McKinsey1999, Guo2009JLTP}. After a prescribed drift time $t_d$, a $5$-ns sheet-shaped imaging pulse at $905$~nm excites the molecules to emit fluorescence at $640$~nm, which is recorded by an intensified sCMOS camera. Because the fluorescence and optical collection efficiencies are limited, we repeat this tagging--drift--imaging sequence typically at $500$~Hz, accumulating 2000 snapshots to obtain a tracer-line image with sufficient signal-to-noise ratio. This ensemble averaging precludes measurement of shot-to-shot turbulent velocity fluctuations, but provides a robust determination of the mean velocity profile, which is the focus of the present study. A typical MTV measurement therefore requires about $4$~s of nearly steady flow. Owing to the finite bellows stroke, this limits the attainable mean flow velocity $U$ and hence restricts the present velocity-profile measurements to $Re\lesssim1.3\times10^6$. At the temperatures studied, the He$_2^*$ molecules are entrained by the normal fluid \cite{Guo2009,Gao2015,Marakov2015}, so the initially straight tracer line (Fig.~\ref{Fig1}d) is displaced and deformed by the local normal-fluid velocity. Figure~\ref{Fig1}e shows representative tracer-line images at two different drift times superimposed, revealing increasing streamwise displacement away from the wall.

\begin{figure}[!t]
	\centering
	\includegraphics[width=1\columnwidth]{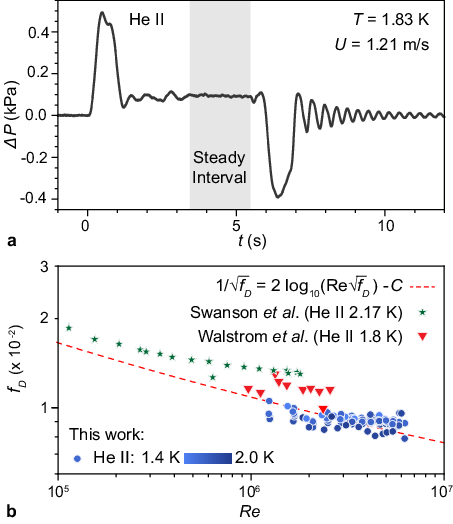}
	\caption{\textbf{Pressure-drop characterization in He II pipe flow.}
	\textbf{a} Differential-pressure trace measured in a representative He II run. The indicated steady interval is used to determine the mean pressure drop $\Delta P$.
	\textbf{b} Darcy friction factor $f_D$ obtained from the measured $\Delta P$ as a function of Reynolds number $Re=UD_h/\nu$. Previous He II measurements by Walstrom \textit{et al.}~\cite{Walstrom1988} and Swanson \textit{et al.}~\cite{Swanson2000} are also shown. The dashed curve denotes a fit to our He II data using the Prandtl--von K\'arm\'an smooth-pipe correlation~\cite{McKeon2004Friction} with an adjustable parameter $C$, yielding $C=0.48\pm0.05$.}
	\label{Fig2}
\end{figure}

The pressure drop is measured independently of the optical velocimetry. As shown in Fig.~\ref{Fig1}a, two pressure taps separated by $L_p=1.83$~m are connected to a Validyne DP10-20 differential-pressure transducer calibrated in liquid helium over $1.5$--$4.2$~K~\cite{Sanavandi2020}. Following the transient oscillatory response at flow onset (Fig.~\ref{Fig2}a), a steady interval is established during which the mean pressure drop $\Delta P$ can be evaluated. At low flow speeds, $\Delta P$ becomes comparable to the background fluctuations, limiting reliable pressure-drop measurements in He II to $Re\gtrsim10^6$. Using the measured $U$ and $\Delta P$, we obtain the Darcy friction factor $f_D$, shown in Fig.~\ref{Fig2}b as a function of $Re=UD_h/\nu$, where $\nu=\mu_n/\rho_{\mathrm{He}}$ with $\mu_n$ being the dynamic viscosity of the normal fluid~\cite{DonnellyBarenghi1998}. Our $f_D$ values are consistent with previous forced-flow measurements in He II~\cite{Walstrom1988,Swanson2000}, despite the differences in their flow pipes. To establish a robust $f_D(Re)$ relation, we fit our He II data to the Prandtl--von K\'arm\'an smooth-pipe correlation, $1/\sqrt{f_D}=2\log_{10}\left(Re\sqrt{f_D}\right)-C$, with $C$ treated as an adjustable parameter. The fitted value, $C=0.48\pm0.05$, is slightly below the classical value of $0.8$~\cite{McKeon2004Friction} (see Methods). Because the MTV and pressure-drop measurements only partially overlap in Reynolds number, $\tau_w$ is obtained directly from $\Delta P$ when reliable $\Delta P$ data are available and from the fitted $f_D(Re)$ relation at lower $Re$. The uncertainty associated with the inferred $f_D$ is incorporated into the Monte Carlo analysis of the wall-law parameters, as detailed in Methods.


\begin{figure*}[t]
\centering
\includegraphics[width=1\linewidth]{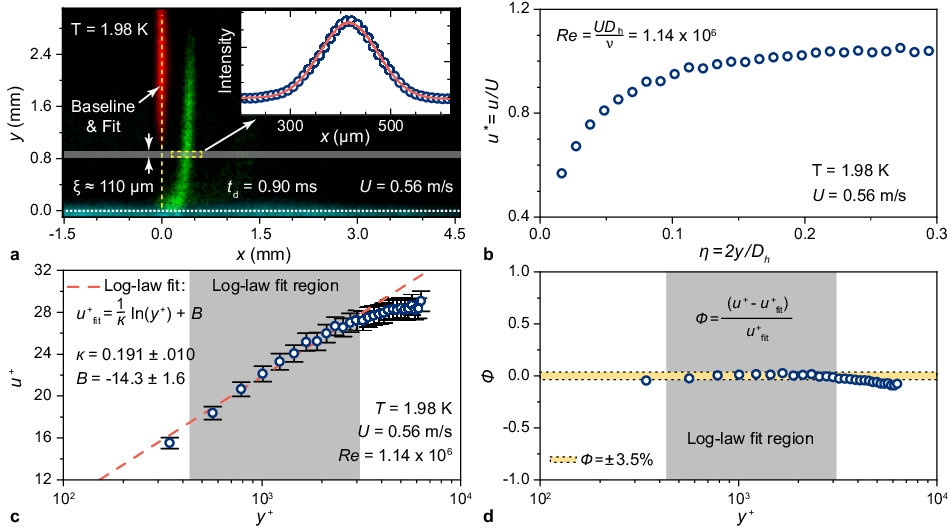}
\caption{\textbf{A nonclassical logarithmic velocity profile in He II.}
\textbf{a} Representative MTV analysis for a He II flow at $T=1.98$~K and $U=0.56$~m/s. The drifted tracer-line image, formed by accumulating 2000 single-shot exposures, is divided into successive wall-normal bins of width $\xi\simeq110~\mu$m. Within each bin, the tracer-line displacement is measured relative to the undeformed baseline. The inset shows the streamwise fluorescence-intensity profile for a representative bin and the Gaussian fit used to determine the tracer-line centre.
\textbf{b} Extracted normal-fluid velocity profile in outer coordinates, $U^*=u/U$ versus $\eta=2y/D_h$.
\textbf{c} The same profile expressed in wall variables, $u^+$ versus $y^+$. The error bars denote the $1\sigma$ experimental uncertainties propagated through the Monte Carlo analysis (see Methods). The shaded region denotes the Reynolds-number-based wall-law region used for the logarithmic fit, and the dashed line shows $u^+_{\mathrm{fit}}=(1/\kappa)\ln y^++B$ using the fitted values of $\kappa$ and $B$.
\textbf{d} Normalized residual $\Phi=(u^+-u^+_{\rm fit})/u^+_{\rm fit}$ of the logarithmic fit. The horizontal dotted lines mark $\Phi=\pm3.5\%$.}
\label{Fig3}
\end{figure*}

\section*{Observation of a nonclassical wall law}
With the MTV images and $\tau_{w}$, we now examine the near-wall scaling in He II. Figure~\ref{Fig3} presents the velocity-profile analysis for a representative He II flow at $T=1.98$~K and $U=0.56$~m/s ($Re=1.14\times10^6$). As illustrated in Fig.~\ref{Fig3}a, the drifted tracer-line image is divided into successive wall-normal bins of width $\xi\simeq110~\mu$m, chosen to be comparable to the tracer-line thickness and hence the effective spatial resolution of the measurement. Within each bin, the streamwise fluorescence-intensity profile is fitted with a Gaussian to determine the tracer-line centre (see Methods), as shown in the inset. The obtained displacement $\Delta x(y)$ relative to the undeformed baseline is then divided by the drift time $t_d$ to obtain the streamwise normal-fluid velocity, $u(y)=\Delta x(y)/t_d$.

The resulting mean velocity profile is shown in Fig.~\ref{Fig3}b in dimensionless form, with $u^*=u/U$ and $\eta=2y/D_h$. The velocity rises rapidly away from the wall and approaches the bulk-flow value toward the pipe interior, revealing a strongly sheared near-wall region. To examine whether this profile exhibits a logarithmic wall law, we recast the data in wall variables using the friction velocity $u_{\tau}=(\tau_w/\rho_{\mathrm{He}})^{1/2}$. As shown in Fig.~\ref{Fig3}c, $u^+$ varies nearly linearly with $\ln y^+$ over an extended range. To identify the wall-law region more objectively, we adopt the Reynolds-number-dependent criteria developed for classical high-Reynolds-number pipe flows~\cite{Marusic2013}. Specifically, the recommended wall-law region is given by $y^+_{\min}<y^+<y^+_{\max}$, with $y^+_{\min}=\max(100,3\sqrt{Re_\tau})$ and $y^+_{\max}=\max(150,0.15Re_\tau)$, where $Re_\tau=u_\tau(D_h/2)/\nu$ is the friction Reynolds number. For the representative case, these criteria render $430\leq y^+\leq3100$ (Fig.~\ref{Fig3}c). Applying the Monte Carlo fitting procedure over this region yields $\kappa=0.191\pm0.010$ and $B=-14.3\pm1.6$, where the uncertainties account for both propagated experimental uncertainties and residual scatter about the logarithmic fit (see Methods).

Our image analysis is restricted to wall distances $y\gtrsim150~\mu\mathrm{m}$, both because of the finite spatial resolution set by the tracer-line thickness and to avoid wall-proximity effects (see methods). The measurements therefore do not extend sufficiently close to the wall to resolve the viscous and buffer layers~\cite{Pope2000}. Nonetheless, this does not affect the present analysis on the logarithmic region, which is clearly resolved. The quality of the logarithmic description is quantified by the normalized residual $\Phi=(u^+-u^+_{\rm fit})/u^+_{\rm fit}$, as shown in Fig.~\ref{Fig3}d. Throughout the identified wall-law region, the residuals remain within $\Phi=\pm3.5\%$. At larger $y^+$, the increasing deviation from the log-law fit marks the transition toward the outer flow.

\begin{figure}[t]
\centering
\includegraphics[width=1\linewidth]{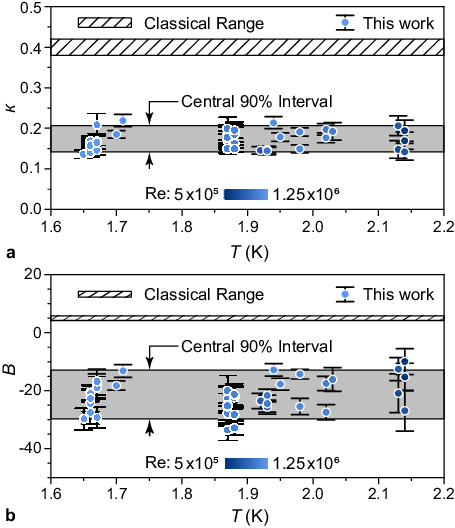}
\caption{\textbf{Wall-law parameters for turbulent He II pipe flow.} \textbf{a} Effective K\'arm\'an constant $\kappa$ obtained from the Monte Carlo analysis of the logarithmic velocity profiles as a function of temperature. The data cluster around $\overline{\kappa}=0.171\pm0.022$, well below the classical smooth-wall range $\kappa\simeq0.38$--$0.42$ indicated by the shaded band~\cite{NagibChauhan2008,McKeon2004Mean,Marusic2013,Bailey2014,Furuichi2015,Furuichi2018}. \textbf{b} Corresponding fitted additive constant $B$. The shaded band denotes the range $B\simeq4.4$--$5.6$ reported for classical high-$Re$ smooth-wall pipe flows~\cite{McKeon2004Mean,Furuichi2015,Furuichi2018}. The error bars in both panels denote the combined $1\sigma$ uncertainties, including experimental uncertainties propagated through the Monte Carlo analysis and residual scatter about the logarithmic fits (see Methods).}
\label{Fig4}
\end{figure}

The representative profile in Fig.~\ref{Fig3} is not an isolated case. Across multiple experimental runs spanning $T=1.65$--$2.14$~K and $Re=5.19\times10^5$--$1.26\times10^6$, all measured He II normal-fluid velocity profiles exhibit a clear logarithmic behavior near the wall. The corresponding wall-law parameters extracted from these profiles are summarized in Fig.~\ref{Fig4} and listed in Supplementary Table~2. The effective K\'arm\'an constants determined from the Monte Carlo analysis cluster around a mean value of $\overline{\kappa}=0.171\pm0.022$, with a central 90\% interval of 0.142--0.207. This mean value is far below the classical smooth-wall range of $\kappa\simeq0.38$--$0.42$ established in high-$Re$ pipe flows~\cite{NagibChauhan2008,McKeon2004Mean,Marusic2013,Bailey2014,Furuichi2015,Furuichi2018}. The corresponding fitted intercepts $B$ are also substantially shifted from the values $4.4$--$5.6$ reported for classical smooth-wall pipe flows~\cite{McKeon2004Mean,Furuichi2015,Furuichi2018}, consistent with the much steeper logarithmic slope $1/\kappa$. Despite appreciable run-to-run scatter, neither $\kappa$ nor $B$ exhibits a systematic dependence on temperature or Reynolds number. These results demonstrate that the modified logarithmic wall law is a robust feature of turbulent He II pipe flow over the explored parameter range.

\section*{Implications of He II wall turbulence}
The invariance of the wall-law parameters with temperature is noteworthy. From $T=1.65$ to $2.14$~K, the normal-fluid fraction $\rho_n/\rho_{\mathrm{He}}$ increases from about $0.19$ to $0.84$~\cite{DonnellyBarenghi1998}, yet neither $\kappa$ nor $B$ exhibits a systematic temperature dependence. In turbulent He II, mutual friction can strongly couple the normal fluid and superfluid on scales larger than the mean intervortex spacing, so that the two fluids develop nearly locked velocity fields, and behave much like a single fluid at a coarse-grained level~\cite{Morris2008,Roche2009,Salort2010,Biferale2018,Tang2020PRFluids}. The dissipative response of this coupled flow then arises from both normal-fluid viscosity and mutual friction.~\cite{Gao2018Dissipation}. This picture may explain why the near-wall structure changes little as the two-fluid composition varies. At the same time, mutual friction exerts an additional force on the normal fluid and modifies its near-wall momentum balance, possibly rendering the nonclassical $\kappa$ and $B$. Therefore, the weak temperature dependence and the modified wall-law parameters may represent two manifestations of the same vortex-mediated coupling. Establishing this connection quantitatively will require future theoretical and numerical work.

As $T$ approaches and crosses the lambda transition into normal liquid helium (He I), one would expect the modified He II wall law to cross over to the classical wall law. We attempted same measurements across this transition. The pressure-drop measurements in saturated He I indeed yielded friction factors that agree well with the classical Prandtl--von K\'arm\'an smooth-pipe correlation (see Supplementary Fig.~3). However, the MTV measurements turn out to be impractical close to the lambda point and in He I, because vapor bubbles interfere with the formation and imaging of well-defined tracer lines. This limitation does not arise from laser absorption by the liquid itself: liquid helium is essentially transparent at the tagging and imaging wavelengths, and only a minute fraction of the $^4$He atoms is ionized to form He$_2^*$ tracers~\cite{Gao2015}, so direct laser heating of the liquid is negligible. Laser absorption by the coated optical windows is small because their transmission exceeds $99\%$. Nevertheless, sufficiently close to the lambda point, and particularly in He I, even these small heat inputs, together with parasitic radiative heating, can nucleate vapor bubbles in the viewing region and thereby prevent reliable MTV measurements. Only at lower temperatures in He II does its exceptional heat-transfer capability suppress such bubble formation~\cite{VanSciver2012} and hence enable MTV.

In summary, our results show that the modified wall law persists over a broad range of He II temperatures. These findings extend a central concept of classical wall turbulence into the quantum-fluid regime and provide a new framework for understanding wall-bounded quantum turbulence. They also have direct practical relevance, as He II is widely used in large-scale cryogenic systems~\cite{VanSciver1992,Claudet1990,Lebrun2000}. The friction behavior and wall-law scaling established here should therefore provide useful guidance for modeling and designing forced-flow He II systems.

\section*{References}
\bibliographystyle{naturemag}
\bibliography{BoundaryMTV_ref}

\section*{Methods}\label{SecVI}
\subsection*{Pressure measurements and friction-factor analysis}

The Validyne DP10-20 differential-pressure transducer was calibrated directly in liquid helium over $1.5$--$4.2$~K following the procedure established for the LHeFT~\cite{Sanavandi2020}. During calibration, one side of the sensing diaphragm was connected to high vacuum while a controlled helium-gas pressure was applied to the other. The transducer output was read through a Validyne CD19A carrier demodulator and recorded as a function of the imposed pressure difference. The response remained linear over a range well exceeding that encountered in the present measurements. A fit to the He II calibration data near $1.8$~K yielded a sensitivity of $\mathrm{d}P/\mathrm{d}V=4.0\pm0.1~\mathrm{kPa/V}$, which was used to convert the measured voltage signals to $\Delta P$.

The bulk flow speed was determined independently from the bellows motion using a Micro-Epsilon ILD-1220 laser displacement sensor with a resolution of $1~\mu\mathrm{m}$ and an acquisition rate of $1$~kHz. The bellows velocity $v_{\rm b}(t)$ was obtained from local linear fits to the measured displacement, and the corresponding pipe velocity was calculated as $U(t)=(A_{\rm b}/D^2)v_{\rm b}(t)$. The bellows geometry produces an essentially constant volumetric displacement rate over the measurement interval~\cite{Sanavandi2020}. For each flow condition, $U$ and $\Delta P$ were evaluated over the quasi-steady interval following the initial flow transient. The temperature-dependent total helium density $\rho_{\rm He}$ and normal-fluid dynamic viscosity $\mu_n$ were evaluated at the measured bath temperature using the saturated-vapor-pressure properties compiled by Donnelly and Barenghi~\cite{DonnellyBarenghi1998}.

The measured He II friction factors were fitted to the Prandtl--von K\'arm\'an relation described in the main text. For MTV measurements at Reynolds numbers below the range of reliable pressure-drop measurements, $f_D$ was evaluated from this fitted relation at the measured $Re$. The uncertainty associated with $f_D$ was accounted for in the Monte Carlo analysis of the wall-law constants. The complete He II pressure-drop dataset and corresponding friction factors are compiled in Supplementary Table~1.

\subsection*{MTV measurements and image analysis}
The femtosecond tagging and $905$-nm imaging laser beams were spatially overlapped in the measurement region using the optical setup developed for the LHeFT~\cite{Sanavandi2020}. The femtosecond beam was focused to a waist radius of about $\omega_0\simeq40~\mu$m, producing a tracer line approximately $100~\mu$m thick over a $\sim5$-mm Rayleigh range. The imaging beam was shaped into a sheet with a streamwise width of about $2.2$~mm and a transverse thickness of $1.1$~mm, providing sufficient coverage of the displaced tracer line. The optical setup and the beam geometry are shown in details in Supplementary Fig.~1.

For each experimental run, the wall position and spatial calibration were established from a boundary-reference image acquired with the same optical and camera configuration. Because the tightly focused femtosecond tagging beam can damage an optical window near its focus and degrade its transparency, a circular aperture of radius $\sim200~\mu\mathrm{m}$ was fabricated in the lower window to allow the tagging beam to pass through. The aperture opens into a short extension tube indium-sealed to the underside of the flow pipe and closed by a coated optical window positioned well outside the focal region, thereby maintaining a sealed flow boundary while avoiding laser damage. The aperture radius is of order $10^2$ wall units under the present conditions. Previous measurements over surfaces containing arrays of perforations with comparable wall-unit dimensions showed that the logarithmic velocity profile retains the universal slope and hence $\kappa$ value~\cite{Wilkinson1983}. In our case, only a single isolated aperture is present, and the velocity data used for the logarithmic fits are measured at wall-normal distances well beyond the aperture scale, further minimizing its influence on the extracted wall-law slope. The femtosecond beam was aligned normal to the lower wall using a co-aligned $635$-nm reference beam. The beam-steering mirrors were adjusted until the reflection from the lower wall retraced and overlapped the incident beam path, thereby establishing normal incidence; the alignment was further verified from zero-flow tracer images. The upper and lower boundaries were determined by Gaussian fits to their wall-normal intensity profiles; their fitted separation was assigned to the known pipe width $D=20$~mm to determine the run-specific pixel scale, while the fitted lower boundary defined $y=0$.

The drift time was defined as $t_d=t_{\rm IMG}-t_{\rm FS}$, where $t_{\rm FS}$ and $t_{\rm IMG}$ are the arrival times of the tagging and imaging pulses, respectively. The sCMOS intensifier gate was synchronized with the imaging pulse and restricted to $8$--$10~\mu$s to suppress background light (see Supplementary Fig.~2). Repeated tagging--drift--imaging cycles acquired under identical conditions were accumulated to form each tracer-line image, while a zero-drift image acquired under the same conditions provided the undeformed reference.

To extract the velocity profile, each drifted tracer line was divided into successive wall-normal bins centered at $y=y_i$, with a bin width $\xi\simeq110~\mu\mathrm{m}$ comparable to the tracer-line thickness. Within each bin, the fluorescence intensity was integrated over its wall-normal extent to obtain the streamwise profile $I_i(x)$, which was fitted with a Gaussian to determine the local tracer-line center $x_i$. Its displacement $\Delta x_i$ relative to the undeformed reference then gives the normal-fluid velocity as $u(y_i)=\Delta x_i/t_d$.

\subsection*{Monte Carlo uncertainty analysis and logarithmic fitting}
The uncertainties in the wall-scaled velocity profiles and the fitted logarithmic-law parameters were evaluated using a Monte Carlo uncertainty-propagation procedure. This approach accounts for both point-specific measurement uncertainties and uncertainties in quantities that are common to an entire velocity profile and therefore introduce correlated variations among the data points. The principal uncertainty sources include the fitted tracer-line positions, the wall location and spatial calibration, and the friction velocity $u_\tau$. The thermophysical properties of He II used in the analysis were treated as accurately known at each experimental temperature.

For each experimental run, the spatial calibration was determined as $\Omega=D/(Y_1-Y_0)$, where $Y_0$ and $Y_1$ are the fitted pixel positions of the lower and upper boundaries, respectively. The local velocity and wall-normal coordinate were calculated as $u_i=\Omega(X_{d,i}-X_{b,i})/t_d$ and $y_i=\Omega(Y_i-Y_0)$, where $X_{d,i}$ and $X_{b,i}$ are the fitted pixel positions of the centers of the drifted and undeformed tracer-line segments. The uncertainties in $X_{d,i}$ and $X_{b,i}$ were obtained from the corresponding Gaussian fits, while those in $Y_0$ and $Y_1$ were obtained from the boundary determination. In each Monte Carlo realization, these measured quantities were randomly sampled from Gaussian distributions centered on their measured values with standard deviations given by their estimated $1\sigma$ uncertainties. The point-specific tracer-line positions were sampled independently, whereas $Y_0$ and $Y_1$, which are common to the entire profile, were sampled once per realization and used to recalculate the common pixel scale $\Omega$ and all $u_i$ and $y_i$ values. This procedure naturally retains the correlations introduced by the common wall position and spatial calibration.

The friction velocity was likewise recalculated in each Monte Carlo realization from the experimentally determined quantities entering the pressure-drop analysis. For conditions with directly resolved $\Delta P$, $u_\tau=[\Delta P D_h/(4\rho_{\rm He}L_p)]^{1/2}$, with the measured pressure drop and its calibration uncertainty included in the resampling. For lower-$Re$ conditions where $f_D$ was inferred from the fitted friction-factor relation, $u_\tau=U\sqrt{f_D/8}$, with the experimental uncertainties associated with $U$ and the inferred $f_D$ included. Quantities that are common to a given run were sampled once per realization and applied consistently to all points in that profile. The wall-scaled variables were then recalculated as $u_i^+=u_i/u_\tau$ and $y_i^+=y_i u_\tau/\nu$. In this way, the correlated effect of $u_\tau$ on both $u^+$ and $y^+$ is retained explicitly.

For each Monte Carlo realization, the same set of data points identified within the logarithmic interval from the nominal profile was fitted to $u_i^+=m\ln y_i^++B$, with $\kappa=1/m$. Each fit yielded values of $\kappa_r$ and $B_r$, together with their regression uncertainties $\sigma_{\kappa,r}^{\mathrm{fit}}$ and $\sigma_{B,r}^{\mathrm{fit}}$ arising from the residual scatter of the data about the fitted logarithmic law. Specifically, $\sigma_{\kappa,r}^{\mathrm{fit}}=\sigma_{m,r}/m_r^2$, where $\sigma_{m,r}$ is the regression uncertainty of the fitted slope, while $\sigma_{B,r}^{\mathrm{fit}}$ is obtained directly from the regression. The complete resampling and fitting procedure was repeated for $10^5$ realizations until the distributions of $\kappa$ and $B$ were statistically converged.

The reported $\kappa$ and $B$ are the means of these Monte Carlo distributions. Their uncertainties, $\sigma_\kappa$ and $\sigma_B$, account for both the propagation of experimental uncertainties and the scatter of the data about the logarithmic fit. The former is quantified by the standard deviations $\sigma_{\kappa,\mathrm{MC}}$ and $\sigma_{B,\mathrm{MC}}$ of the fitted values across the Monte Carlo ensemble, while the latter is represented by the root-mean-square regression uncertainties $\sigma_{\kappa,\mathrm{fit}}=[\langle(\sigma_{\kappa,r}^{\mathrm{fit}})^2\rangle_r]^{1/2}$ and $\sigma_{B,\mathrm{fit}}=[\langle(\sigma_{B,r}^{\mathrm{fit}})^2\rangle_r]^{1/2}$. The final $1\sigma$ uncertainties are therefore $\sigma_\kappa=(\sigma_{\kappa,\mathrm{MC}}^2+\sigma_{\kappa,\mathrm{fit}}^2)^{1/2}$ and $\sigma_B=(\sigma_{B,\mathrm{MC}}^2+\sigma_{B,\mathrm{fit}}^2)^{1/2}$. The Monte Carlo ensemble alone is used to determine the experimental uncertainties of the wall-scaled coordinates shown in Fig.~\ref{Fig3}c. The run-specific quantities used in the wall normalization and the resulting logarithmic-fit parameters are compiled in Supplementary Table~2.

\section*{Data availability}
\noindent The MTV images and the processed velocity-profile data are available from the corresponding author upon reasonable request.

\section*{Code availability}
\noindent All computer codes used in this study are available from the corresponding author upon reasonable request.

\section*{Acknowledgments}
The authors thank R. Keijzer and S. Bao for assistance with the MTV optical setup, Y. Zhang for developing the LabVIEW data-acquisition system, H. Sanavandi for restoring the LHeFT and contributing to preliminary measurements, and T. Kanai, Y. Qi, and Y. Xing for assistance with the current experiments. The authors also thank L. Cattafesta and C. Meneveau for valuable discussions on wall-law physics. This work was supported by the Gordon and Betty Moore Foundation through Grant GBMF11567. The experiments were conducted at the National High Magnetic Field Laboratory at Florida State University, which is supported by National Science Foundation Cooperative Agreement No. DMR-2128556 and the State of Florida.\\

\section*{Author contributions}
\noindent W.G. conceived and supervised the research; M.H., M.S., and S.I. performed the experiments and analyzed the data; W.G. and M.H. drafted the manuscript; all authors discussed the results and reviewed the final manuscript.

\section*{Competing interests}
\noindent The authors declare no competing interests.

\end{document}